\documentclass[aps,prb,twocolumn,showpacs,superscriptaddress]{revtex4-1}
\usepackage{amsfonts,mathrsfs,amsmath,amsthm,amssymb,bbold}
\usepackage{epsfig}
\usepackage{bm,mathrsfs}
\usepackage{lineno,hyperref}
\usepackage{amsmath}
\usepackage{color}
\usepackage{xcolor,colortbl} 
\usepackage{soul}
\usepackage{afterpage}
\usepackage{capt-of}
\usepackage{makecell}
\usepackage{graphicx}
\setcellgapes{4pt}
\newcolumntype{P}[1]{>{\centering\arraybackslash}p{#1}}

\begin{document}  
\title {\bf Topology analysis for anomalous Hall effect in the non-collinear antiferromagnetic states of \textbf{Mn}$\mathbf{_3}$\textbf{\textit{A}N (\textit{A} = Ni, Cu, Zn, Ga, Ge, Pd, In, Sn, Ir, Pt)}} 

\author{Vu Thi Ngoc Huyen}
\affiliation{Institute of Scientific and Industrial Research, Osaka University, 8-1 Mihogaoka, Ibaraki, Osaka 567-0047, Japan}
\affiliation{Research and Services Division of Materials Data and Integrated System, National Institute for Materials Science, 1-2-1 Sengen, Tsukuba, Ibaraki 305-0047, Japan}
\affiliation{Graduate School of Engineering Science, Osaka University, Toyonaka, Osaka 560-8531, Japan}

\author{Michi-To Suzuki}
\thanks{Electronic address: michito.suzuki@imr.tohoku.ac.jp}
\affiliation{Center for Computational Materials Science, Institute for Materials Research, Tohoku University, Sendai, Miyagi 980-8577, Japan}

\author{Kunihiko Yamauchi}
\affiliation{Institute of Scientific and Industrial Research, Osaka University, 8-1 Mihogaoka, Ibaraki, Osaka 567-0047, Japan}

\author{Tamio Oguchi}
\affiliation{Institute of Scientific and Industrial Research, Osaka University, 8-1 Mihogaoka, Ibaraki, Osaka 567-0047, Japan}
\affiliation{Research and Services Division of Materials Data and Integrated System, National Institute for Materials Science, 1-2-1 Sengen, Tsukuba, Ibaraki 305-0047, Japan}
\date{\today}
\begin{abstract}
We investigate topological features of electronic structures which produce large anomalous Hall effect in the non-collinear antiferromagnetic metallic states of anti-perovskite manganese nitrides by first-principles calculations. We first predict the stable magnetic structures of these compounds to be non-collinear antiferromagnetic structures characterized by either $T_{1g}$ or $T_{2g}$ irreducible representation by evaluating the total energy for all of the magnetic structures classified according to the symmetry and multipole moments. The topology analysis is next performed for the Wannier tight-binding models obtained from the first-principles band structures. Our results reveal the small Berry curvature induced through the coupling between occupied and unoccupied states with the spin-orbit coupling, which is widely spread around the Fermi surface in the Brillouin zone, dominantly contributes after the ${\bm k}$-space integration to the anomalous Hall conductivity, while the local divergent Berry curvature around Weyl points has a rather small contribution to the anomalous Hall conductivity.
\end{abstract}
\maketitle
\section{Introduction}\label{sec1}
Anomalous Hall (AH) effect has been focused on exploring the relation between the topological feature of electronic band structures and its emergence as a macroscopic phenomenon \cite{reviewahe}. Recently, the large AH effect was predicted by the first-principles calculations for non-collinear antiferromagnets with no net magnetization \cite{PRL112,2014nahc,mn3x,2017mn3gesn} and was observed experimentally for the antiferromagnetic (AFM) phases in Mn$_3$Sn and Mn$_3$Ge \cite{2015mn3sn, 2016mn3ge1, 2016mn3ge2}. 
The large AH effect in AFM states has attracted an increasing amount of attention because of the insensitivity against an applied magnetic field and no stray fields interfering with the neighboring cells as well as faster spin dynamics than ferromagnets \cite{2015mn3sn,2016mn3ge1,2016mn3ge2,2017mn3gesn,mn3sn2017ex,nernst2018ex}.
Those findings of the AH effect in the non-collinear AFM states urge us to get a comprehensive understanding of possible AH effect in various magnetic states. 

One of the authors has shown that some antiferromagnetic structures can induce the AH effect by breaking the magnetic symmetry same as that for the ordinary ferromagnetic order, and introduced cluster multipoles to identify the order parameters which induce the AH effect as a natural extension of magnetization in ferromagnets \cite{cmp2017, cmp2018}. 
In this context, anti-perovskite manganese nitrides can be regarded as a new playground to explore the AH effect, since  Mn${_3}A$N ($A$= Ni, Sn) have been found to show non-collinear AFM in the triangular Mn lattice corresponding to irreducible representations $T_{1g}(\Gamma^+_4)$ and $T_{2g}(\Gamma^+_5)$, respectively\cite{1978ex, 2010ex, 2013ex} and there are many analogues with the replaced nonmagnetic elements. 
A recent study on the spin-order dependent AH effect in the non-collinear AFM Mn$_3A$N ($A$= Ga, Zn, Ag, or Ni) also suggested that these compounds are an excellent AFM platform for realizing novel spintronics applications \cite{zhou2019}.
 
The AH effect was suggested mainly arising from large Berry curvature around the Weyl points in Weyl semimetals \cite{claudia2018, topologicalsemimetals}. For metallic ferromagnetic bcc-Fe, Mart\'inez {\it et al.} investigated topological feature related to the AH effect and found the dominant contribution from the Berry curvature distribution across the Fermi sheets with the possible enhanced contribution from the Fermi sheets having the Weyl points very nearby \cite{PRB085138}.
In this paper, we provide the results of systematic analysis for the AH effect in anti-perovskite manganese nitrides Mn${_3}A$N ($A$= Ni, Cu, Zn, Ga, Ge, Pd, In, Sn, Ir, Pt) and discuss the stability, symmetry, and topology aspects of the magnetic structures leading to the AH effect.
In particular, we identify important factors for the large AH effect with the detailed analysis of Weyl points, Berry curvature, and Fermi surfaces, which characterize the topological features of the magnetic systems, by means of first-principles calculations.
We find that the AH effect is dominantly contributed from the Berry curvatures widely spread around the Fermi surfaces induced with the band splitting due to the spin-orbit coupling (SOC) and the contribution from the divergent Berry curvature, for instance, around Weyl points is rather small.

This paper is organized as follows. Section \ref{secsym} shows symmetry analysis related to AH effect in Mn$_3A$N. The method to perform the first-principles calculation is presented in Sec.~\ref{secmethod}. Then results for electronic and topological aspects of the AH conductivity in these compounds are shown in Sec.~\ref{secresults}. We investigate the stable magnetic structures in Sec.~\ref{secresults} A and the AH conductivity in Sec.~\ref{secresults} B. In Sec.~\ref{secresults} C, we show Weyl points can produce divergent peaks of the Berry curvature when they are located just around the Fermi level, but the contribution to the AH effect is nevertheless small. We then discuss the dominant factor that contributes to the AH conductivity in Sec.~\ref{secresults} D. Finally, Sec.~\ref{conclusion} contains a summary of this work.
\section{Symmetry and anomalous Hall effect in M\lowercase{n}$\mathbf{_3}$\textbf{\textit{A}}N} \label{secsym}
\begin{figure} 
\centering
\includegraphics[width=8.6cm]{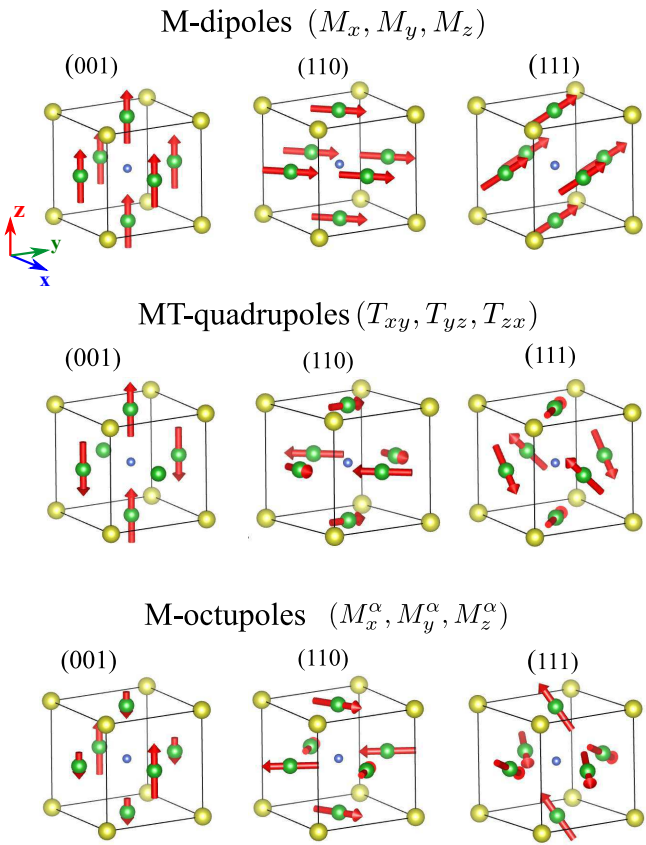} 
\captionof{figure}{Energetically inequivalent magnetic structures of $\mathrm{Mn_3}A\mathrm{N}$ classified according to the multipole moments following Ref.~\onlinecite{cmpgeneration}. The green, yellow, and blue balls indicate Mn, $A$, and N atoms, respectively. Arrows on Mn atoms indicate the magnetic moments.} 
\label{fig:vesta2}
\end{figure}
\begin{table}
\captionof{table}{Classification of the magnetic structures with the ordering vector ${\bm q}$=0 in $\mathrm{Mn_3}A\mathrm{N}$ according to the symmetry-adapted multipole~\cite{cmpgeneration} as well as the irreducible representation (IR), magnetic point group (Mag. PG) with its principal axis (P. axis). The AH conductivity tensors (AHC) that can be finite under the magnetic point groups are also listed, where $\sigma_{110}\equiv \frac{1}{\sqrt{2}}(\sigma_{yz}+\sigma_{zx})$ and $\sigma_{111}\equiv \frac{1}{\sqrt{3}}(\sigma_{yz}+\sigma_{zx}+\sigma_{xy})$.}
\label{tab:orthonormal}
\begin{tabular}{|P{0.9cm}|P{3.4cm}|P{1.6cm}|P{1.0cm}|P{0.8cm}|} 
\hline
 $O_h$-IR& Multipole & Mag. PG & P. axis & AHC\\
\hline
$T_{1g}$& $(M_x,M_y,M_z)= (001)$ & $4/mm'm'$&[100]&$\sigma_{yz}$\\
              &\hspace{2.1cm}$=(010)$
              &$4/mm'm'$&[010]&$\sigma_{zx}$\\
              &\hspace{2.1cm}$=(001)$ &$4/mm'm'$&[001]&$\sigma_{xy}$\\
              &\hspace{2.1cm}$=(110)$ &$m'm'm$&[110]&$\sigma_{110}$\\
              &\hspace{2.1cm}$=(111)$ &$\bar{3}m'$&[111]&$\sigma_{111}$\\
$T_{2g}$& $(T_{yz},T_{zx},T_{xy})= (100)$ &$4'/mm'm$&[100]& None\\
    &\hspace{2.1cm}$=(010)$ &$4'/mm'm$&[010]& None\\
    &\hspace{2.1cm}$=(001)$ &$4'/mm'm$&[001]& None\\
    &\hspace{2.1cm}$=(110)$ & $mm'm$ & [110]& None\\
    &\hspace{2.1cm}$=(111)$ & $\bar{3}m$ &[111]& None\\
$T_{1g}$& $(M_x^{\alpha},M_y^{\alpha},M_y^{\alpha})=(100)$ &$4/mm'm'$&[100]&$\sigma_{yz}$\\
              &\hspace{2.2cm}$=(010)$ &$4/mm'm'$&[010]&$\sigma_{zx}$\\
              &\hspace{2.2cm}$=(001)$ &$4/mm'm'$&[001]&$\sigma_{xy}$\\
              &\hspace{2.2cm}$=(110)$&$m'm'm$ & [110]&$\sigma_{110}$\\
              &\hspace{2.2cm}$=(111)$ &$\bar{3}m'$&[111]&$\sigma_{111}$\\
\hline
\end{tabular}  
\end{table}
Manganese nitrides Mn$_3A$N have the anti-perovskite crystal structure which belongs to the space group $Pm\bar{3}m$ ($O^1_h$, No.~221). We classify the energetically inequivalent magnetic structures with the ordering vector $\bm q= 0$, shown in Fig.~\ref{fig:vesta2}, using the symmetry-adapted multipole magnetic structure bases generated following Ref.~\onlinecite{cmpgeneration}. In Fig.~\ref{fig:vesta2}, the magnetic (M)-dipole structures $(M_x, M_y, M_z)= (001)$, $(110)$, and $(111)$ represent ferromagnetic structures oriented along [001], [110], and [111] directions, respectively. 
The pure antiferromagnetic structures are obtained as the magnetic structures orthogonalized to the M-dipole structures \cite{cmpgeneration} and are, in this compound, obtained as the rank-2 magnetic toroidal multipoles (MT-quadrupoles) and rank-3 M-multipoles (M-octupoles).

Orthogonalized multipoles which belong to $T_{1g}$ and $T_{2g}$ IRs are listed in Table \ref{tab:orthonormal} together with the non-zero AH conductivity tensors. As shown in Table \ref{tab:orthonormal}, the M-octupoles can induce the AH effect since these ordered states break the magnetic symmetry same as those of the M-dipoles \cite{cmp2017}. On the other hand, MT-quadrupoles, which belong to $T_{2g}$ IR, do not induce the AH effect with the magnetic structures shown in Fig.~\ref{fig:vesta2} due to the presence of the magnetic symmetry which forbids the finite AH conductivity as we demonstrate in Sec.~\ref{secresults}.

As discussed in Ref.~\onlinecite{cmp2017}, co-planar magnetic structures induce no AH effect in the absence of SOC in general by the presence of the effective time-reversal symmetry, which is the symmetry of conjunct operation of the time-reversal and global spin rotation. 
The M-dipoles and M-octupoles in Fig.~\ref{fig:vesta2} need SOC to induce the AH effect. In the following section, we proceed to the quantitative evaluation of the AH conductivity for the M-octopole structure based on the results of first-principles calculations considering the SOC.
\section{Method}\label{secmethod}
\begin{figure} 
\centering 
\includegraphics[width=7.9cm]{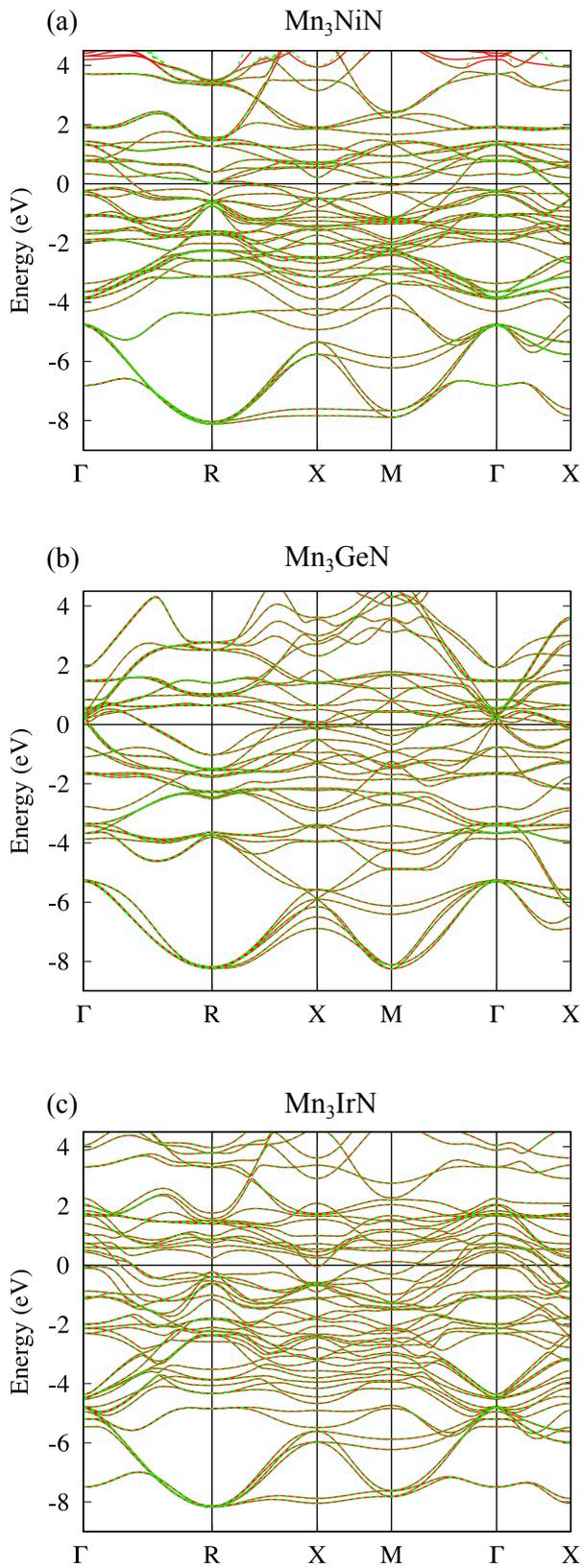}\\
\captionof{figure}{Energy bands from the first-principles calculations (red) and from Wannier interpolation (green) of (a) $\mathrm{Mn_3NiN}$, (b) $\mathrm{Mn_3GeN}$, and (c) $\mathrm{Mn_3IrN}$ along high symmetry points in the first Brillouin zone of a simple cubic shown in Fig.~\ref{fig:bz}.}
\label{fig:bandfitting}
\end{figure}
\begin{figure} 
\centering 
\includegraphics[width=6.0cm]{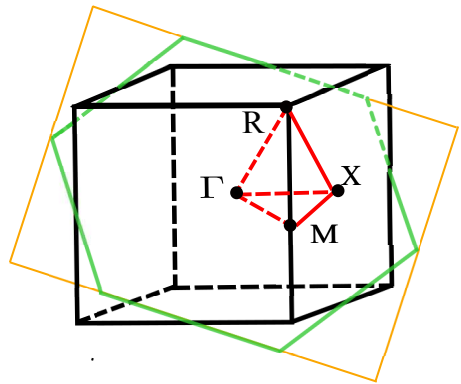}
\captionof{figure}{The first Brillouin zone (black) corresponding to the crystal primitive unit cell with the high symmetry points.The hexagonal plane (green line) shows minimum periodicity in the (111) plane for the simple cubic Brillouin zone with the center point at $\Gamma$. The orange rectangular is the region used to plot the Berry curvature in Fig.~\ref{fig:berry}.}
\label{fig:bz}
\end{figure}
QUANTUM ESPRESSO package \cite{QE} is used to perform first-principles calculations and to evaluate the electronic and magnetic properties of antiperovskite manganese nitrides. Generalized gradient approximation in the parametrization of Perdew, Burke, and Ernzerhof \cite{GGA-PBE} is used for the exchange-correlation functional. The pseudopotentials in the projector augmented-wave method \cite{paw} are generated by PSLIBRARY \cite{pslibrary}. We choose kinetic cut-off energies 100 Ry and 800 Ry for the plane wave basis set and charge density, respectively.

The AH conductivity is calculated by the Brillouin zone integration of the Berry curvature with summation of the one-electron bands below the Fermi level \cite{ahc,ahc2}:
\begin{equation}
\label{equ:ahc}
\begin{split}
\sigma_{\alpha\beta}=-\frac{e^2}{\hbar}\int\frac{d\textbf{\textit{k}}}{(2\pi)^3}\sum_nf_n(\textbf{\textit{k}})\Omega_{n,\alpha\beta}(\textbf{\textit{k}})
\end{split}
\end{equation}
where $n$ is band index, $\alpha,\beta= x,y,z$ ($\alpha\neq\beta$), and $f_n(\textbf{\textit{k}})=\theta(\epsilon_n(\textbf{\textit{k}})-\mu)$ is the occupation factor determined from the eigenvalue of the Bloch states $\epsilon_n(\textbf{\textit{k}})$ and the Fermi energy $\mu$. The Berry curvature is evaluated following the Kubo formula \cite{ahc, ahc4}:
\begin{equation}
\label{equ:berry}
\begin{split}
\Omega_{n,\alpha\beta}(\textbf{\textit{k}})= -2 \mathrm{Im} \sum_{m\neq n}\frac{v_{nm,\alpha}(\textbf{\textit{k}})v_{mn,\beta}(\textbf{\textit{k}})}{[\epsilon_m(\textbf{\textit{k}})-\epsilon_n(\textbf{\textit{k}})]^2}
\end{split}
\end{equation}
where the velocity operator is defined in term of the periodic part $u_{n\textbf{\textit{k}}}$ of the Bloch states:
\begin{equation}
\label{equ:velocity}
\begin{split}
v_{nm,\alpha}(\textbf{\textit{k}})=\frac{1}{\hbar}\left \langle u_{n}(\textbf{\textit{k}})\left|\frac{\partial \hat{H}(\textbf{\textit{k}})}{\partial k_{\alpha}}\right | u_{m}(\textbf{\textit{k}}) \right \rangle
\end{split}
\end{equation}
with $\hat{H}(\textbf{\textit{k}})=e^{-i\textbf{\textit{k.r}}}\hat{H}e^{i\textbf{\textit{k.r}}}$.
The AH conductivity is evaluated by using the realistic tight-binding models obtained from the first-principles band structures \cite{ahc} by Wannier interpolation scheme using Wannier90 \cite{w90}.
Including $s, p, d$ orbitals for Mn and $A$ atoms and $s, p$ orbitals for N atoms, we have obtained the tight-binding models showing almost complete reproducibility of the energy bands for those obtained from the first-principles calculations within the energy interval from the lowest energy of the valence bands to about 4 eV above the Fermi energy for the Mn$_3A$N series, as shown in Fig.~\ref{fig:bandfitting} for $\mathrm{Mn_3GeN}$, $\mathrm{Mn_3PdN}$, and $\mathrm{Mn_3IrN}$.
A ${\bm k}$-mesh 18$\times$18$\times$18 is utilized to sample the first Brillouin zone (BZ) with Methfessel-Paxton smearing width of 0.005 Ry to get the Fermi level. 
The AH conductivity was evaluated with the uniform ${\bm k}$-point mesh of 200$\times$200$\times$200 with the adaptive ${\bm k}$-mesh refinement \cite{adap1,adap2} of 5$\times$5$\times$5 for the absolute values of Berry curvature larger than 100$\mathrm{\AA^2}$.
\section{Results}\label{secresults}
\subsection{Stability of magnetic structure in Mn${\mathbf{_3}}$\textbf{\textit{A}}N}
\begin{table*}
\captionof{table}{Equilibrium lattice constants $a_0\mathrm{(\AA)}$, local magnetic moments $|m\mathrm{_{local}|(\mu_B)}$, and the difference of total energy $\mathrm{\Delta}E$ (meV/f.u.) between (111) magnetic orderings and the
M-octupole ($M_x^{\alpha}$, $M_y^{\alpha}$ $M_z^{\alpha}$)= (111) 
(MO) configurations. The bold values indicate the lowest $\mathrm{\Delta}E$. The M-dipole ($M_x,M_y,M_z$)= (111) and M-T quadrupole ($T_{yz}, T_{zx}, T_{xy}$)= (111) are referred as the FM [111] and the MTQ configuration, respectively.} 
\label{tab:energy}
\begin{center}
\begin{tabular}{|P{0.5cm}|P{1.5cm}|P{0.9cm}|P{1.2cm}|P{1.1cm}|P{1.5cm}|P{7.7cm}|} 
\hline
\multicolumn{6}{|c|}{\textbf{This work}} & \multicolumn{1}{c|}{\textbf{Experiments}}\\
\hline
 $A$  & Config.    & $a_0 ($\AA$)$ & $|m_{\mathrm{local}}|$ $(\mu_B)$ & $|m_{\mathrm{total}}|$ $\mathrm{ (\mu_B)}$ & $\mathrm{\Delta}E$  (meV/f.u.) & Magnetic configurations (temperature) \\
\hline
&FM [111] & 3.827  & 3.12 & 9.35   &  345.5 &  $\bullet$  MO + MTQ ($10K<T< 250K$) \cite{2010ex} \\
\cline{2-6} 
Ni  &MTQ  & 3.832  & 2.99 &  0.0 &  0.04  & $\bullet$ MO + MTQ $(160K <T<266 K)$  \cite{1978ex} \\
\cline{2-6} 
& MO  &  3.832 & 2.99 &  0.0 & \textbf{0} & \\
\hline
& FM [111] & 3.851 & 2.74 & 8.23 & 257.8 &  \\
\cline{2-6} 
Cu & MTQ & 3.853  & 2.87  & 0.0 & \textbf{-7.5}  & $\bullet$ Ferromagnetic in tetragonal  ($T < 150K$) \cite{2001ex} \\ 
\cline{2-6}  
& MO & 3.853  &  2.97& 0.10 & 0 & \\
\hline
& FM [111] & 3.781 & 1.510 & 4.53  & 190.8 & $\bullet$ AFM but not  MTQ ($T < 80K$) \cite{2012ex}\\
\cline{2-6} 
Zn &MTQ & 3.866  & 2.74 & 0.0 & \textbf{-0.4}& $\bullet$ MTQ $(80K <T<170K)$ \cite{2012ex,1978ex}  \\
\cline{2-6}  
& MO & 3.866  &  2.74 & 8.23 & 0 & \\ 
\hline
& FM [111] & 3.757  & 1.07 & 3.23 & 124.8& \\
\cline{2-6} 
Ga& MTQ & 3.865 & 2.61 & 0.00&   \textbf{-0.4} &   $\bullet$ MTQ $(T<298K)$  \cite{1978ex}\\
\cline{2-6} 
& MO & 3.865 & 2.61 & 0.08 & 0& \\
 \hline
& FM [111] & 3.756 & 0.91& 2.73 & 146.3 & \\
\cline{2-6} 
Ge & MTQ  & 3.858 & 2.49  & 0.0  & \textbf{-8.6} & - \\
\cline{2-6} 
& MO & 3.858 & 2.49 &  0.0 & 0&  \\
 \hline
& FM [111]  & 3.949 & 3.21& 9.66 & 474.6 &  \\
\cline{2-6} 
Pd & MTQ  & 3.927 & 3.36  & 0.0  & \textbf{-9.5} &- \\
\cline{2-6} 
& MO & 3.927 & 3.34 &  -0.01 & 0&  \\
\hline
& FM [111] & 3.910  & 1.56 & 4.68    &  329.3  &  $\bullet$ Weak FM+ AFM ($T <175K$) \cite{2012ex}  \\
\cline{2-6} 
In & MTQ  &  3.989 &  2.61 &  0.0  &    74.6 &  $\bullet$ AFM ($175K<T< 300K$) \cite{2012ex} \\
\cline{2-6} 
& MO & 3.989 & 2.91  & 0.05 & \textbf{0}& \\ 
 \hline
&FM [111] &3.882 &1.193 & 3.58 & 236.7 &$\bullet$ Complex magnetic ordering $(T< 237 K)$ \cite{1977ex} \\
\cline{2-6} 
Sn & MTQ & 3.851 & 2.01 & 0.0   & 215.6 &   $\bullet$  MO and MTQ $(237K<T<357K)$ \cite{1978ex,1977ex}\\
\cline{2-6} 
& MO& 3.982 & 2.75   &  0.0 & \textbf{0}   & \\ 
\cline{2-6} 
 \hline
&FM [111] &3.870&2.94&8.81& 807.8 &\\
\cline{2-6} 
Ir & MTQ  &  3.863 & 2.77 &  0.00  &  \textbf{-3.0} &  - \\
\cline{2-6} 
& MO  & 3.863 & 2.77 & 0.06 & 0.0 & \\
  \hline
& FM [111]  & 3.949 & 3.25& 9.66 & 483.0 &\\
\cline{2-6} 
Pt & MTQ  & 3.927 & 3.23  & 0.0  & \textbf{-6.7} & - \\
\cline{2-6} 
& MO  & 3.927 & 3.23 & - 0.05 & 0& \\
\hline 
\end{tabular} 
\end{center}
\end{table*}
\begin{figure} 
\centering 
\includegraphics[width=8.0cm]{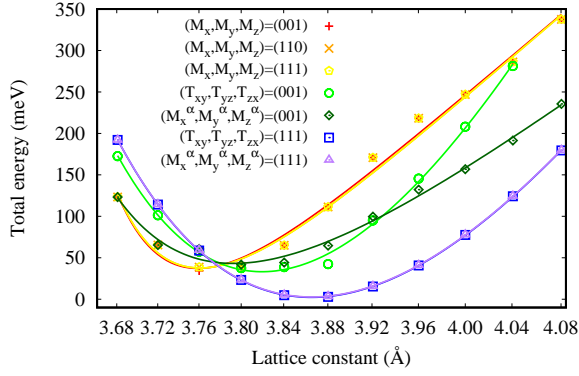}
\captionof{figure}{Total energy as the function of lattice constants for different magnetic configurations in $\mathrm{Mn_3GaN}$. The equilibrium total energy of the $(M_x^{\alpha},M_y^{\alpha},M_z^{\alpha})= (111)$ magnetic structure is chosen as the origin of total energy. The values are fitted to Birch-Murnaghan's equation of state \cite{Murnaghan} by the least square method.}
\label{fig:ev}
\end{figure}
We first consider the stability of magnetic structures in Mn$_3A$N by comparing total energies calculated by the first-principles approach. The optimization of lattice constants for each magnetic structure in Mn$_3A$N are performed by calculating lattice constant dependence of the total energy as shown for Mn$_3$GaN in Fig.~\ref{fig:ev}. The optimized lattice constants agree with previous experimental values \cite{2014ex,1981ex}. It is shown that either ($T_{yz}$, $T_{zx}$, $T_{xy}$)= (111) or ($M_x^{\alpha}$, $M_y^{\alpha}$, $M_z^{\alpha}$)= (111) is obtained as the stable magnetic structure in Mn${_3}A$N.
We hereafter focus on these (111) non-collinear AFM structures, MTQ and MO and refer the magnetic structures of ($T_{yz}$, $T_{zx}$, $T_{xy}$)= (111) and of ($M_x^{\alpha}$, $M_y^{\alpha}$, $M_z^{\alpha}$)= (111) as MT-quadrupole (MTQ) and M-octupole (MO), respectively, following the multipole characterization of the magnetic structure proposed in Ref.~\onlinecite{cmpgeneration}. The total energies for ferromagnetic, MTQ, and MO magnetic structures are listed in Table \ref{tab:energy} with the relative energy from the MO magnetic structure, \textit{i.e.} $\Delta{E}= E-E_{\rm MO}$, for the series of Mn$_3A$N.

Table \ref{tab:energy} shows that Mn${_3}$\textit{A}N with \textit{A} = Ni, In, Sn prefer the MO configuration, and those with the other $A$ atoms prefer the MTQ configuration, having the MO magnetic structure as the secondary stable solution. The energy differences between the MO and MTQ magnetic structures are small for most of the Mn$_3A$N compounds. Mn$_3$NiN shows only tiny energy difference of 0.04 meV/f.u., which explain the experimentally reported possible coexistence of the MO and MTO phases~\cite{2010ex}. 
On the other hand, we may expect that Mn${_3}$InN  and Mn${_3}$SnN are stabilized to the MO phase with $\Delta E (\rm MTQ - MO) \sim$ 74.6 and 215.6 meV/f.u. and active for the AH effect. 
The presence of weak ferromagnetism in AFM states observed for Mn$_3$InN \cite{2012ex} implies that the observed AFM structure is the MO structure since the MO and ferromagnetic structures belong to the same magnetic symmetry and can coexist in the magnetic phase. 
In the followings, we will focus on the AH effect in the MO magnetic structure, which is the first or secondary stable solution for all of Mn$_3A$N and can induce the AH effect. 
\subsection{Anomalous Hall conductivity} 
\begin{figure} 
\centering    
\includegraphics[width=8.6cm]{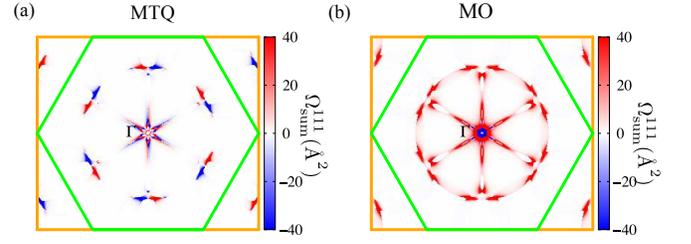}\\
\captionof{figure}{The [111] Berry curvature component after taking band summation, $\Omega^{111}_{\mathrm{sum}}$(\AA$^2$)$\equiv \frac{1}{\sqrt{3}}(\Omega_{yz,\mathrm{sum}}+\Omega_{zx,\mathrm{sum}}+\Omega_{xy,\mathrm{sum}})$, on (111) plane centered at $\Gamma$, shown in Fig.~\ref{fig:bz}, for Mn$_3$GeN with (a) the MTQ and (b) the MO configuration, respectively.}
\label{fig:berry}
\end{figure} 
\begin{table}
\captionof{table}{Calculated AH conductivity, $\sigma_{111}$, for the MO magnetic configuration in Mn$_3A$N compounds.}
\label{tab:ahc}
\begin{center}
\begin{tabular}{|P{1.2cm}|P{1.5cm}|P{1.6cm}|P{3.2cm}|} 
\hline 
         & Mn$_3A$N   &This work & References \\
\cline{2-4} 
       & &  & -301 \cite{zhou2019}$^{*}$\\
       & Mn$_3$NiN & 375.7 &-294.5 ($\sigma_{xy}=$-170) \cite{expahcmn3nin}$^{*}$\\
       &&&  225.2 ($\sigma_{xy}=130$) \cite{ahcmn3gan}$^{**}$\\
       \cline{2-4}
       & Mn$_3$CuN & -287.7 & - \\
       \cline{2-4}
$\sigma_{111}$ & Mn$_3$ZnN & 350.5 & -232 \cite{zhou2019}$^{*}$ \\
       \cline{2-4}
       & Mn$_3$GaN & 96.3 & -99 \cite{zhou2019}$^{*}$ \\
(S/cm)       &&& 69.3 ($\sigma_{xy}$=40) \cite{ahcmn3gan}$^{**}$ \\
       \cline{2-4}
       & Mn$_3$GeN & -624.5 & -  \\
       \cline{2-4}
       & Mn$_3$PdN & 252.6 & -  \\
       \cline{2-4}
       & Mn$_3$InN & 34.6 & -  \\
       \cline{2-4}
       & Mn$_3$SnN & -128.0 & 230.4 ($\sigma_{xy}=133$) \cite{ahcmn3gan}$^{**}$\\
       \cline{2-4}
       & Mn$_3$IrN & -575.3 &  - \\
       \cline{2-4}
       & Mn$_3$PtN & 799.9 & -  \\
\hline
\multicolumn{4}{l}{$^{*}$\footnotesize{Magnetic moments assumed to calculate the AH conductivity}} \\
\multicolumn{4}{l}{\footnotesize{are opposite to those of this work.}} \\
\multicolumn{4}{l}{$^{**}$\footnotesize{The sign of the AH conductivity listed in Ref.~\onlinecite{ahcmn3gan} is uncertain.}} \\
\end{tabular}  
\end{center}
\end{table}
We have calculated the AH conductivity, $\sigma_{111}\equiv \frac{1}{\sqrt{3}}(\sigma_{yz}+\sigma_{zx}+\sigma_{xy}$), for the magnetic structures shown in Fig.~\ref{fig:vesta2} and listed the values in Table \ref{tab:ahc}.
Note that the conductivity ($\sigma_{yz}$, $\sigma_{zx}$, $\sigma_{zy}$) has the transformation property for the magnetic point group same as that for the magnetization ($M_x$, $M_y$, $M_z$)~\cite{cmp2017}, and the time-reversal counterparts of the magnetic structures hold the opposite sign to the AH conductivity.
Some of Mn$_3A$N materials show the large AH conductivities in the non-collinear AFM magnetic structure as the same order of the AH conductivity calculated for the ferromagnetic states such as Fe (750 S/cm) \cite{adap1,ahc} and Co (480 S/cm) \cite{wang2007}. The AH conductivity values for the non-collinear antiferromagnet $\mathrm{Mn_3Ir}$, which shows  the same magnetic alignment on Mn atoms in Mn$_3A$N, is also evaluated in this work as 233.8 S/cm and in good agreement with the previous work (218 S/cm) \cite{PRL112}.
Some of the AH conductivities theoretically predicted in these compounds are the same order in this work as listed in Table \ref{tab:ahc}. The difference in its value may come from the details of first-principles calculations such as adopting of lattice constants from experiments or from optimization procedures. 
The AH conductivity was recently reported for thin films of Mn$_3$NiN as $|\sigma_{xy}|=$ 15 S/cm at 150K under no external magnetic field~\cite{expahcmn3nin}, which is one order smaller than the theoretical prediction. The large difference with the experiment and theoretical prediction can be addressed to the possible mixing of the MO and the MTQ magnetic structures as discussed in Ref.~\onlinecite{expahcmn3nin}.
 
Figure \ref{fig:berry} shows distribution of the Berry curvature component after taking band summation, 
$\Omega^{111}_{\mathrm{sum}}({\bm k})\equiv \frac{1}{\sqrt{3}}(\Omega_{yz,\mathrm{sum}}({\bm k})+\Omega_{zx,\mathrm{sum}}({\bm k})+\Omega_{xy,\mathrm{sum}}({\bm k}))$ with $\Omega_{\alpha\beta,\mathrm{sum}}({\bm k})=\sum_nf_n({\bm k})\Omega_{n,\alpha\beta}({\bm k})$, on the (111) plane shown in Fig.~\ref{fig:bz} for the MO and MTQ magnetic structures. The MO and MTQ magnetic structures belong to the magnetic point groups $\bar{3}m'$ and $\bar{3}m$, respectively, and the Berry curvature distribution keeps the three-fold rotation symmetry on the (111) plane. In contrast to the MO magnetic structure, the MTQ magnetic structure cancels out the Berry curvature on the (111) plane with BZ integration due to the mirror symmetry with the vertical mirror planes and leads to no AH conductivity for the magnetic structure. 
%
\subsection{Topology analysis}
\begin{figure*} 
\includegraphics[width=16.0cm]{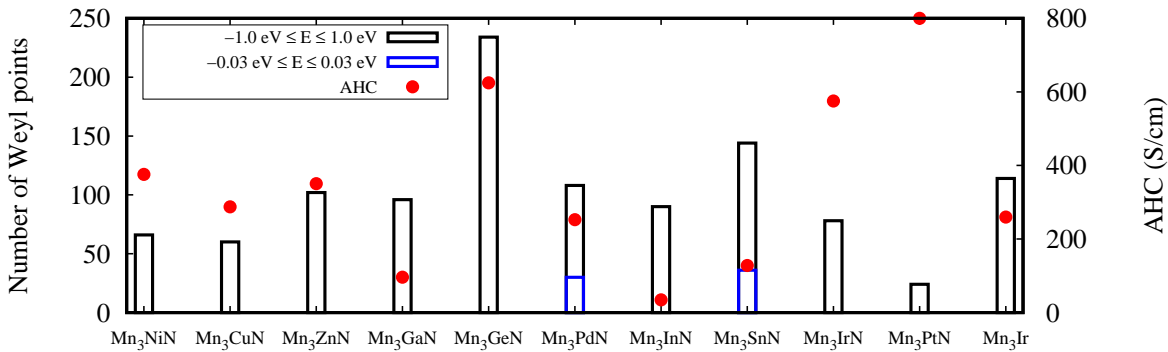}
\captionof{figure}{Number of Weyl points around the Fermi level (black boxes and blue boxes) with the calculated AH conductivity (red dots) for the series of Mn$_3A$N.}
\label{fig:numweyl}
\end{figure*} 
\begin{figure} 
\centering      
\hspace{0.0cm} 
\includegraphics[width=8.5cm]{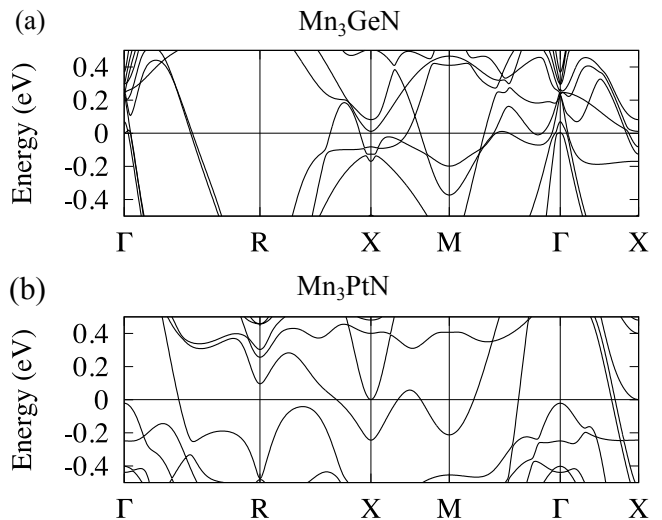}
\captionof{figure}{Band structure in (a) Mn$_3$GeN and (b) Mn$_3$PtN along high symmetry lines in the first Brillouin zone of a simple cubic shown in Fig.~\ref{fig:bz}.}
\label{fig:ptband}
\end{figure} 
\begin{figure} 
\centering      
\hspace{0.0cm} 
\includegraphics[width=8.6cm]{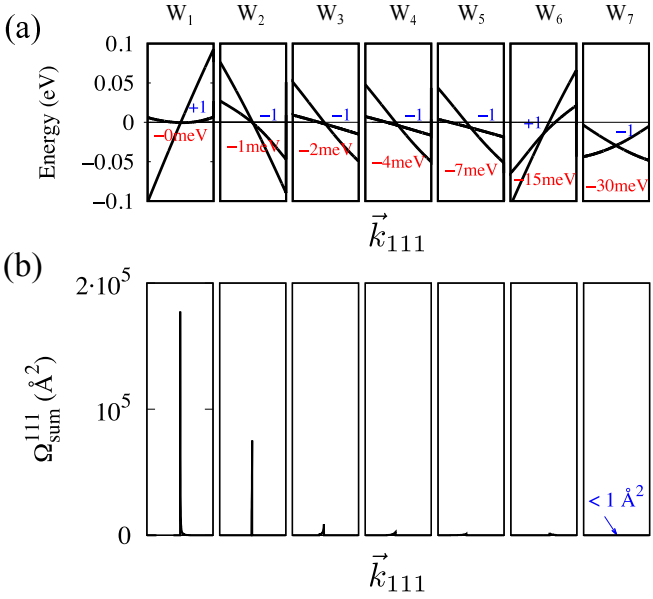}
\captionof{figure}{(a) Band structure and (b) Berry curvature along the [111] direction, $\vec{k}_{111}$, having Weyl points near Fermi energy that produce the positive Berry curvature after taking band summation in Mn$_3$SnN. Each panel shows an interval 0.109 (\AA$^{-1}$) along $\vec{k}_{111}$ with Weyl point at the middle of the line. The relative energies with respect to the Weyl points are written in red, the blue number +1 and -1 indicate the chiralities of the Weyl points. The value ``$- 0$ meV'' indicates the Weyl point within the energy range of -1 meV $<$ E $<$ 0 meV. The coordinates of these Weyl points in the reciprocal space from left to right are W$_1$ = (-0.06, -0.34, -0.34), W$_{2}$ = (-0.04, 0.34, 0.34), W$_{3}$ = (-0.05, 0.44, -0.16), W$_{4}$ = (-0.05, -0.16, 0.44), W$_{5}$ = (-0.16, -0.05, 0.44), W$_{6}$ = (-0.34, -0.34, 0.03) and  W$_{7}$ = (-0.15, 0.47, 0.05), respectively.}
\label{fig:snbandweyl}
\end{figure} 
\begin{figure} 
\centering
\includegraphics[width=8.0cm]{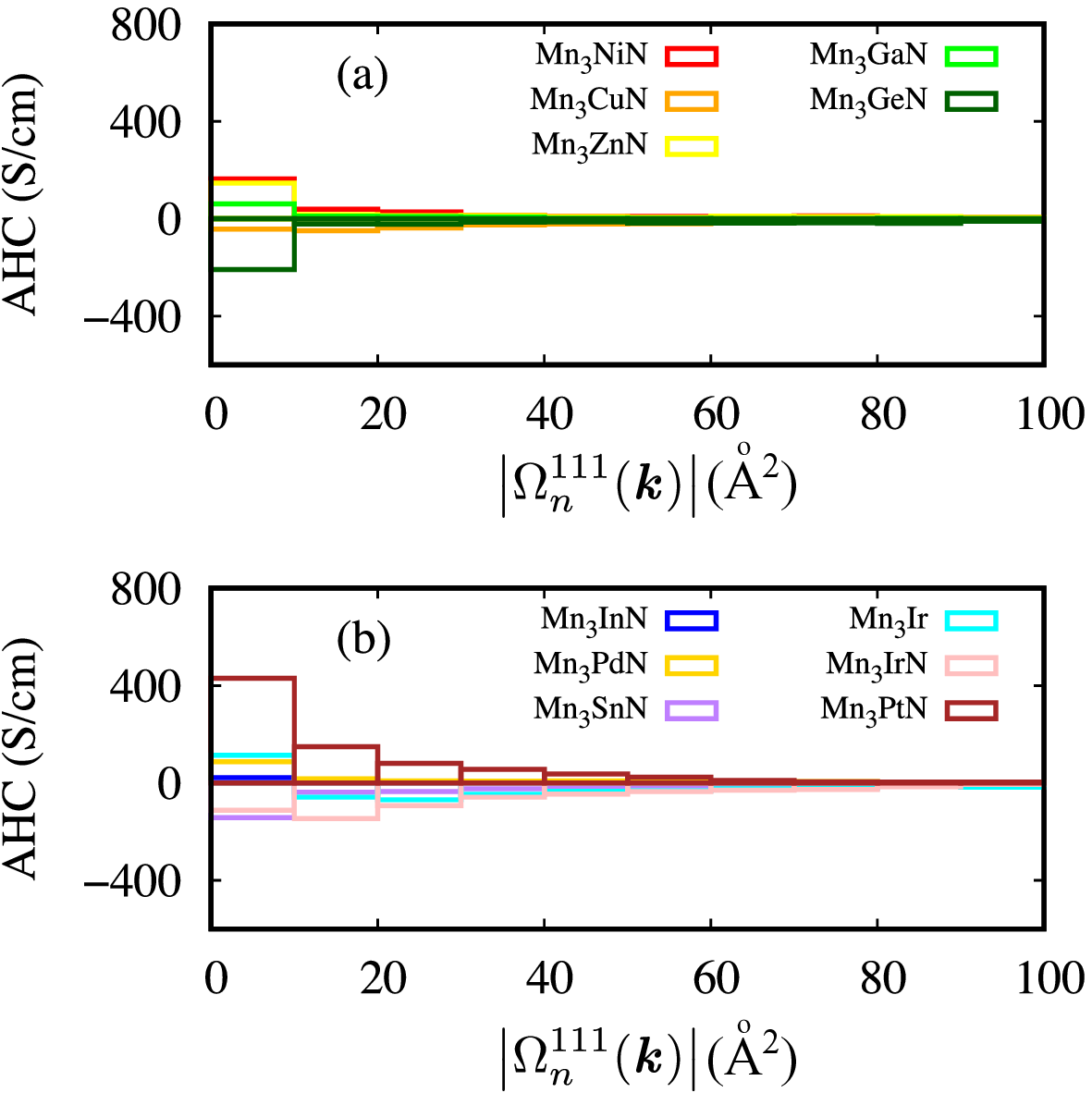}
\captionof{figure}{The bar chart showing contribution of the Berry curvature to the resultant AH conductivity of Mn$_3A$N with the $A$ elements having (a) small and (b) large SOC. The horizontal axis is the absolute intensity of the Berry curvature. The contribution is also shown for Mn$_3$Ir, which shows the same magnetic alignment on Mn atoms in Mn$_3A$N.}
\label{fig:contributionabs}
\end{figure}
In Weyl semimetal, it has been often suggested that the Berry curvature around the Weyl points dominantly contribute to the AH effect in the local ${\bm k}$-space regions\cite{claudia2018,topologicalsemimetals}.
For metallic magnets, Mart\'inez \textit{et al.} suggested that the Fermi sheets with Weyl points very nearby tend to contribute more to the AH conductivity than other Fermi sheets farther from Weyl points by investigating  ferromagnetic bcc Fe \cite{PRB085138}. 
In this section, we investigate the Berry curvature, Weyl points which characterize the topological aspects of the magnetic structures, and their roles in the resultant AH effect for the AFM states in Mn$_3A$N.

We determined Weyl points by examining chirality for possible energy crossing points. The converged number of Weyl points in the BZ is obtained by increasing ${\bm k}$-point mesh in the first BZ to search the crossing points, and the chirality is calculated from the Berry flux coming out of a small sphere $S$ surrounding each Weyl point, \textit{i.e.} $\frac{1}{2\pi}\oint_{S}dS\widehat{\mathbf{n}}.\mathbf{\Omega}_n(\mathbf{k})$ \cite{PRB085138}. 
Figure \ref{fig:numweyl} shows the number of Weyl points around the Fermi level, which are presented in the BZ with the calculated AH conductivity for the series of Mn$_3A$N.
It is shown that there are several Weyl points within the energy range -1.0 eV $<$ E $<$ 1.0 eV in all of the investigated compounds, but only Mn$_3$SnN and Mn$_3$PdN have the Weyl points within $\pm$30 meV around the Fermi level.
Figure \ref{fig:ptband} shows the band structures of Mn$_3$GeN, which shows the maximum number of Weyl points in the energy range -1.0 eV $<$ E $<$ 1.0 eV among the compounds calculated in Fig.~\ref{fig:numweyl}, and Mn$_3$PtN, which shows the minimum number of Weyl points, along high symmetry lines. The difference in the number of Weyl points appears qualitatively as the difference in the complexity of the energy bands around the Fermi level.
Figure \ref{fig:numweyl} displays no strong correlation between the number of Weyl points and the size of the AH conductivity. For instance, Mn$_3$PtN shows the largest AH conductivity for the smallest number of Weyl points around the Fermi level among these compounds.

To investigate the contribution of the Berry curvature around Weyl points to the AH conductivity, we pick up the some Weyl points around the Fermi level in Mn$_3$SnN, which has the maximum number of Weyl points in the energy range -0.03 eV $<$ E $<$ 0.03 eV, and show the band structures around the Weyl points (Fig.~\ref{fig:snbandweyl} (a)) with the resulting Berry curvature after taking the band summation (Fig.~\ref{fig:snbandweyl} (b)). 
Figure \ref{fig:snbandweyl} shows that the Berry curvature around Weyl points contributes to producing the sharp peaks of the band summation of the Berry curvature when the Weyl points are located near the Fermi level within the energy range of 1 meV while the Weyl points located at the energy more than 1 meV below the Fermi energy do not produce finite contribution of the Berry curvature after taking band summation since the crossing bands are both occupied. The $\Omega_{\textrm{sum}}({\bm k})$ enhanced around the Weyl points close to the Fermi level is consistent with the large contributions to the AH conductivity of the Fermi sheets with the Weyl points very nearby as discussed in Ref.~\onlinecite{PRB085138}. The detailed analysis for the contribution of Fermi surfaces to the AH conductivity in Mn$_3A$N is left for future work.

Figure \ref{fig:contributionabs} shows the contribution of the Berry curvature, classified according to its value of $\mid \Omega _n^{111} ({\bm k})\mid$ in the first BZ, where $\Omega_{n}^{111}({\bm k})\equiv \frac{1}{\sqrt{3}}(\Omega_{yz,n}({\bm k})+\Omega_{zx,n}({\bm k})+\Omega_{xy,n}({\bm k}))$ is the [111] Berry curvature component of band $n$ at each $\bm k$ point, to the resultant AH conductivity, $\sigma_{111}$.
Figure \ref{fig:contributionabs} shows the Berry curvature with small value dominantly contribute to the AH conductivity and the contribution rapidly decreases as the value becomes larger. The plot clearly shows that the contribution of the divergent Berry curvature to the AH conductivity is quite small in these AFM states even for the compounds with several Weyl points around the Fermi level leading to the divergent Berry curvature summation at the local ${\bm k}$-region.

We further evaluate the contribution of the divergent Berry curvature around the Weyl points to the AH conductivity for Mn$_3$SnN, which has many Weyl points close to the Fermi level as shown in Fig. \ref{fig:numweyl}, by calculating the ${\bm k}$-integral in Eq.~(\ref{equ:ahc}) within the cubes set around each Weyl point in BZ. Decreasing the size of the cubes, we obtain the converged values of the contribution to the AH conductivity in Mn$_3$SnN around seven percent.
The small contribution of the local divergent Berry curvature to the resultant AH conductivity can be understood from the divergent $\Omega_{\textrm{sum}}^{111}({\bm k})$ region too small to produce a large contribution to the AH conductivity or, otherwise, from canceling it out with the other contribution that has the opposite sign of the Berry curvature at different ${\bm k}$ points in BZ.
%
\subsection{Berry curvature and spin-orbit coupling effect}
\begin{figure} 
\centering      
\includegraphics[width=8.0cm]{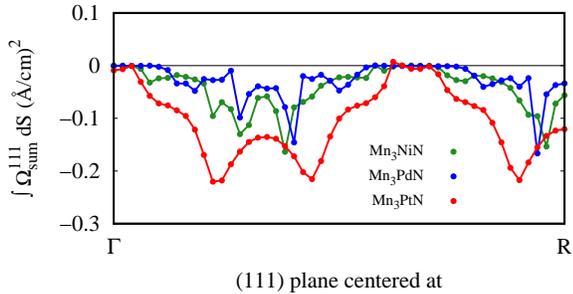} \\
\captionof{figure}{The Berry curvature integrated on the (111) hexagonal area as shown in Fig.~\ref{fig:bz} with its center changing from $\Gamma$ to R for Mn$_3$\textit{A}N (\textit{A}= Ni, Pd, Pt).}
\label{fig:ahciter}
\end{figure} 
\begin{figure*} 
\centering   
\includegraphics[width=16cm]{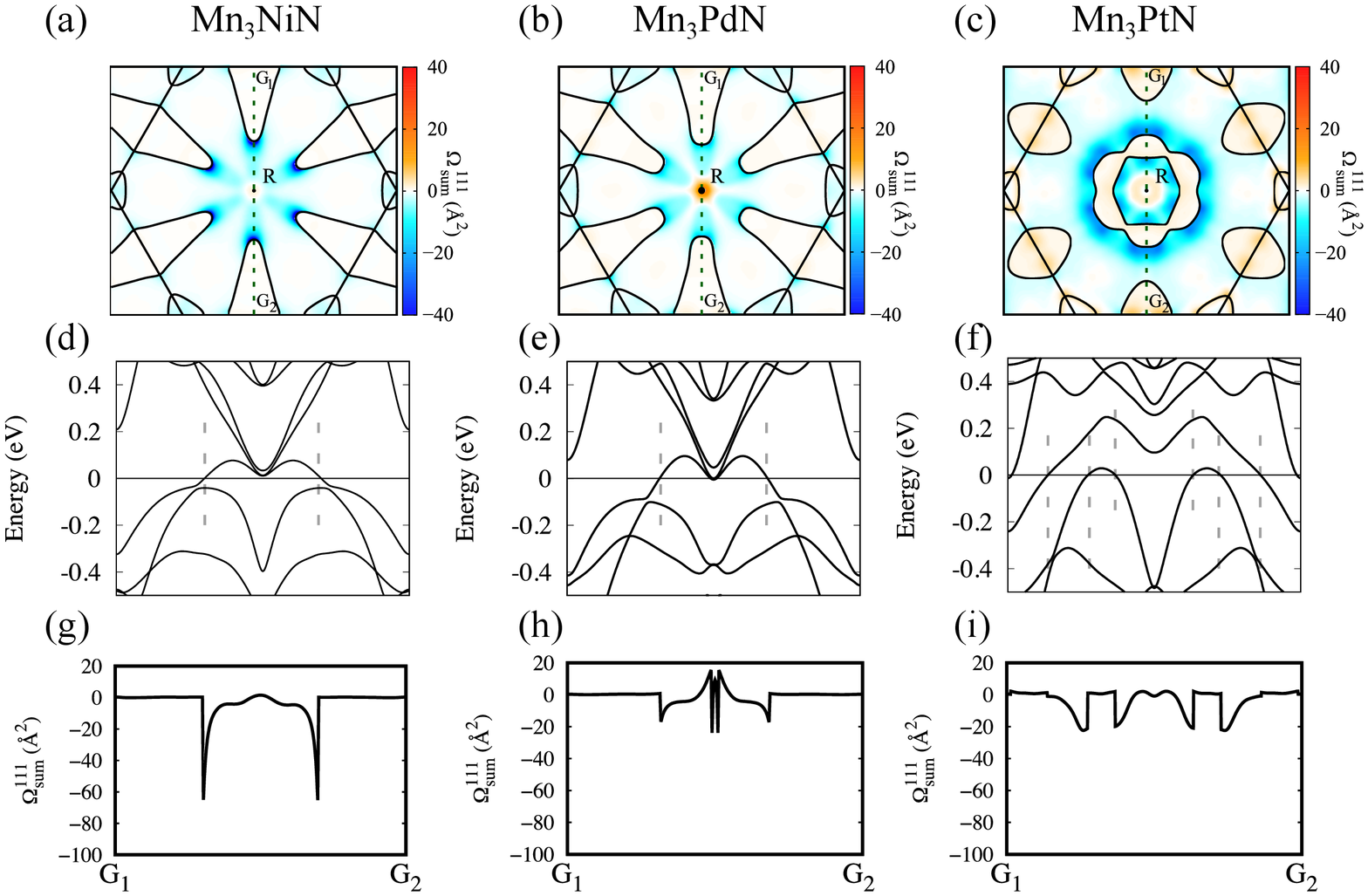}\
 \captionof{figure}{(a, b, c) Distribution of the Berry curvature after taking band summation, $\Omega^{111}_{\rm sum}$, and Fermi surfaces on the BZ plane as shown in Fig.~\ref{fig:bz} with its center point of R. (d, e, f) The band structure around Fermi energy and (g, h, i) Berry curvature $\Omega^{111}_{\rm sum}$ on the $G_1$-$G_2$ line shown in (a, b, c), respectively.}
\label{fig:nipdpt050}
\end{figure*}
We here investigate the electronic structure, Berry curvature, and AH conductivity in the Mn$_3A$N with $A$ = Ni, Pd, and Pt which belong to the same group in the periodic table and are expected to have similar electronic valence states except for the effect of SOC coupling for the purpose to discuss the topological feature which enhance the AH conductivity. 
Figure \ref{fig:ahciter} shows the Berry curvature integrated on the hexagonal plane with the minimum periodicity in the (111) plane, as shown in Fig.~\ref{fig:bz}, moving the center point of the hexagonal plane from $\Gamma$ to R for the three compounds. As shown in Fig.~\ref{fig:nipdpt050}, the integrated Berry curvature shows similar dependency for the (111) plane, starting from the almost zero value for the plane including $\Gamma$ to the negative finite values for the one including R, for these compounds.
The Berry curvature after taking band sum is shown for the (111) plane including R point in the upper panel of Fig.~\ref{fig:nipdpt050}, exhibiting the region with sizable Berry curvature spread around the Fermi surfaces, which we hereinafter call active area of the Berry curvature.

Mn$_3$NiN and Mn$_3$PdN show similar values of the AH conductivity through all of the different (111) planes in Fig.~\ref{fig:ahciter}.
This reflects the similarity of the band structures as shown in Fig.~\ref{fig:nipdpt050} (d) and (e), which result in the similar Fermi surfaces and Berry curvature distribution shown in Fig.~\ref{fig:nipdpt050} (a) and (b). On the other hand, the small difference of the electronic structure can modify the local structure of the Berry curvature distribution as shown in Fig.~{fig:nipdpt050} (g) and (h). 
As shown in Fig.~\ref{fig:nipdpt050} (d) and (g), two sharp negative peaks of the Berry curvature in Mn$_3$NiN come from the two small gaps around the Fermi level. The SOC of Pd, relatively larger than that of Ni, increases those gaps and lower the top peaks for Mn$_3$PdN compared to those for Mn$_3$NiN through the denominator of Eq.~(\ref{equ:berry}), making the possible contribution to the AH conductivity smaller than that for Mn$_3$NiN.
Meanwhile, Mn$_3$PtN exhibits larger active area of the Berry curvature than those for Mn$_3$NiN and Mn$_3$PdN in its absolute value as shown in Fig.~\ref{fig:nipdpt050} (c).
The enhancement of the Berry curvature over BZ for Mn$_3$PtN, which can be seen in Fig.~\ref{fig:ahciter}, is thus associated with the enlarged active area of the Berry curvature through the large SOC of Pt in Mn$_3$PtN and leads to the largest AH conductivity in the calculations among the three compounds.
The enhancements in the cross term of the velocity matrix in Eq.~(\ref{equ:berry}) through SOC for the states around the Fermi surface take place in a broad region of BZ, possibly contributing to the obtained large AH conductivity in the AFM Mn$_3A$N compounds.
%
\section{Conclusions}\label{conclusion}
In summary, we have investigated the stable magnetic structures, the AH effect, and the topology related to the AH effect in anti-perovskite manganese nitrides Mn$_3A$N. Their MO non-collinear AFM states, which are the most or second stable magnetic structures whose magnetic symmetry allows to induce the AH effect, exhibit the AH conductivities comparable to those in ferromagnetic states of Fe and Co in size. 
We have shown that the Berry curvature spread around the Fermi surfaces in the broad BZ region, coming from the band splitting due to the SOC dominantly contribute to the AH conductivity, while the locally divergent Berry curvature produces only a small contribution to the AH conductivity after considering the band summation and BZ integral in Eq.~(\ref{equ:ahc}). It opens a viewpoint for a relation between topology and macroscopic phenomena in non-collinear AFM. Our study might also motivate and guide further various exciting researches in associating with topology and AFM spintronic applications.
%
\section*{Acknowledgement}
We thank F. Kuroda and Y. Yanagi for helpful comments and discussion. This work was supported by the Materials Research by Information Integration Initiative (MI$^{\mathrm{2}}$I) of the National Institute for Materials Science (NIMS), and the International Scientific industrial research (ISIR), Osaka University, JSPS KAKENHI Grants No. JP18H04227, JP15K17713, JP15H05883 (J-Physics), JP17H06154, JP18H04230, JST- PRESTO and CREST No. JPMJCR18T1, Japan Science and Technology Agency.
%
\bibliography{basename of .bib file}

%
\end{document}